\documentclass[10pt,a4paper]{article}
\usepackage[english]{babel}
\usepackage{amsmath,amsfonts,amssymb,bm}\interdisplaylinepenalty=2500
\usepackage{longtable}

 \mathchardef\epsilon="0122   \mathchardef\varepsilon="010F
 \mathchardef\theta="123      \mathchardef\vartheta="0112
 \mathchardef\rho="125        \mathchardef\varrho="011A
 \mathchardef\phi="127        \mathchardef\varphi="011E
\ifx\undefined\degrees\def\degrees{\ensuremath{^{\circ}}}\fi
\ifx\undefined\celsius\def\celsius{\ensuremath{^{\circ}\mathrm{C}}}\fi
\ifx\undefined\unit\def\unit#1{\ensuremath{\mathrm{\,#1}}}\fi
\ifx\undefined\micro\def\micro{\ensuremath{\mu}}\fi
\ifx\undefined\sups\def\sups#1{\ensuremath{^{\mathrm{#1}}}}\fi
\ifx\undefined\subs\def\subs#1{\ensuremath{_{\mathrm{#1}}}}\fi
\ifx\undefined\ohm\def\ohm{\ensuremath{\mathrm{\Omega}}}\fi
\def\req#1{(\ref{#1})}
\ifx\pdfoutput\undefined
  \usepackage{graphicx,color}   % for including graphics
  \usepackage[nativepdf,backref,bookmarks]{hyperref} % dvips
  \usepackage{rotating}
\else
  \usepackage[backref,bookmarks]{hyperref}  %% PDFLATEX
  \usepackage[pdftex]{graphicx,color}     % for including graphics
  \usepackage[pdftex]{rotating}              %  goes after graphicx!!!
  \usepackage{epstopdf}
\fi
\graphicspath{{figs/}}
\def\normalsizescale{0.707}
\def\TINYscale{0.459}
\def\PrintGraphicFileNeme{0}
\newcommand{\namedgraphics}[2]{%     % #1=scale #2=file
	\parbox{\columnwidth}{%
	\ifnum\PrintGraphicFileNeme>0\rotatebox{90}{~\ttfamily\scriptsize#2}\fi%
	\hspace*{\fill}\includegraphics[scale=#1]{#2}\hspace*{\fill}}}
\newcommand{\includefig}[2]{%     % COMBINED LATEX-PS/PDF FIGURES
	\parbox{\columnwidth}{%
	 \ifnum\PrintGraphicFileNeme>0\rotatebox{90}{~\ttfamily\scriptsize#2}\fi%
	\hspace*{\fill}\scalebox{#1}{%  %              % scale
	\ifx\pdfoutput\undefined\input{\includefigpath#2.pstex_t}
	\else\input{\includefigpath#2.pdftex_t}\fi}\hspace*{\fill}}}
\newcommand{\comment}[1]{\footnote{#1}}
\renewcommand{\comment}[1]{}

\newcommand{\ToDay}{Sep.\,2006}
\newcommand{\TODAY}{September 17, 2006}
\title{The effect of AM noise on correlation phase noise measurements}
\author{Enrico Rubiola and Rodolphe Boudot\\
\small web page \texttt{http://rubiola.org}
\\[4em]\includegraphics[width=0.35\textwidth]{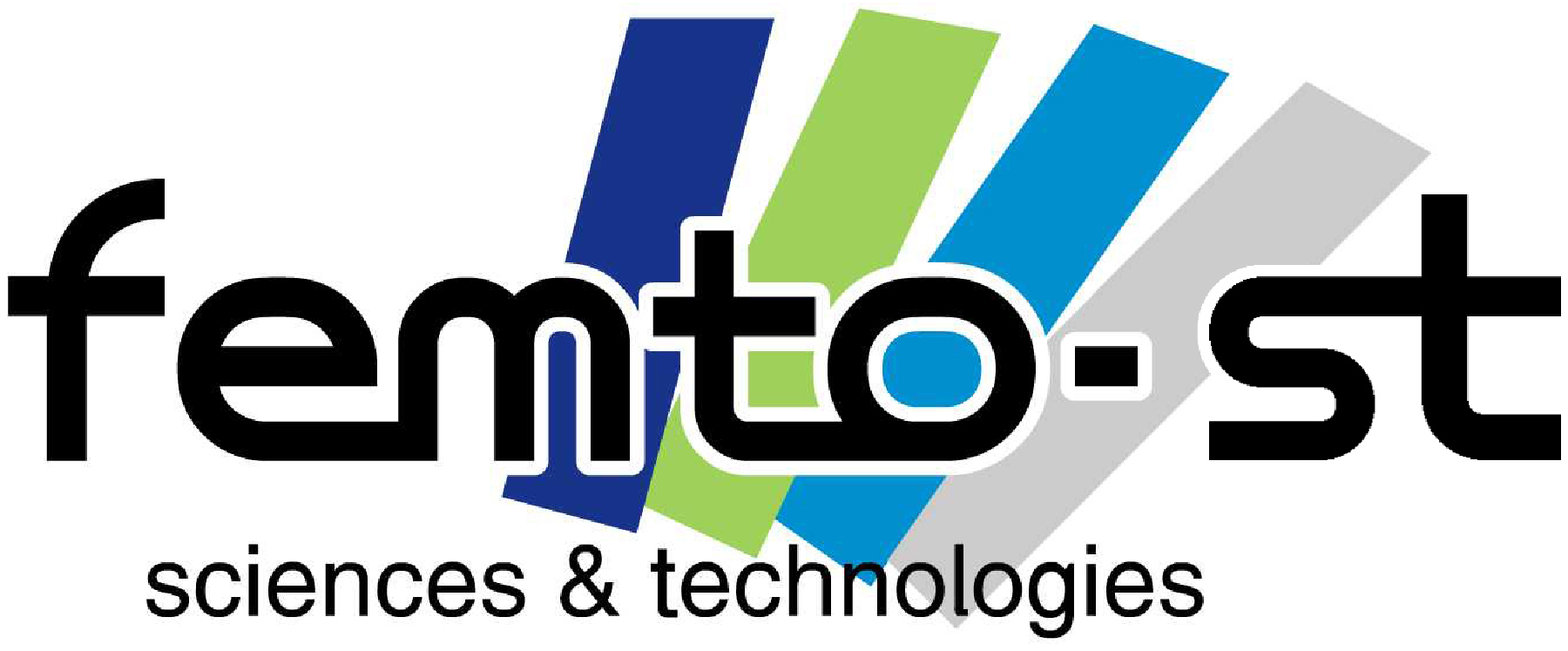}\\[0.5em]
\small FEMTO-ST Institute\\[-0.5ex]
\small CNRS and Universit\'e de Franche Comt\'e, 
\small Besan\c{c}on, France\\[1.5em]}
\date{\small\TODAY}
\begin{document}
\maketitle

\begin{abstract}
We analyze the phase-noise measurement methods in which correlation and averaging is used to reject the background noise of the instrument.  All the known methods make use of a mixer, used either as a saturated phase detector or as a linear synchronous detector.
Unfortunately, AM noise is taken in through the power-to-dc-offset conversion mechanism that results from the mixer asymmetry.   
The measurement of some mixers indicates that the unwanted amplitude-to-voltage gain is of the order of 5--50 mV, which is 12--35 dB lower than the phase-to-voltage gain of the mixer.  In addition, the trick of setting the mixer at a sweet point---off the quadrature condition---where the sensitivity to AM nulls, works only with microwave mixers.  The HF-VHF mixers have not this sweet point.  Moreover, we prove that if the AM noise comes from the oscillator under test, it can not be rejected by correlation.  At least not with the schemes currently used.  An example shows that at some critical frequencies the unwanted effect of AM noise is of the same order---if not greater---than the phase noise.  Thus, experimental mistakes are around the corner. 
\end{abstract}

\clearpage

\begin{center}
\addcontentsline{toc}{section}{Symbol list}
\begin{longtable}{ll}\hline
\multicolumn{2}{l}{\textbf{\large\rule[-1ex]{0pt}{3.5ex}Symbol list}}\\\hline
\rule{0pt}{2.5ex}%
 $\left<\,\right>_m$ 
		& average on $m$ realizations\\
$a(t)$, $b(t)$	& single-channel random signals\\
$c(t)$	& random signal common to the two channels\\
$\mathbb{E}\{\cdot\}$ & statistical expectation\\
$f$		& Fourier frequency (near dc)\\
$g$		& voltage gain (thus, the power gain is $g^2$)\\
$h(t)$	& impulse response of a (linear) system\\
$k_l$	& LO amplitude-to-voltage gain $k_l=v_o/\alpha_\text{LO}$
		    [Eq.~\req{eqn:amx-real-mixer-output-2}]\\
		& in the measurement of oscillators (schemes B, and C)\\
$k_{lr}$	& LO+RF amplitude-to-voltage gain $k_l=v_o/\alpha_\text{LO}$ 
		    [Eq.~\req{eqn:amx-real-mixer-output-1}]\\
		& in the measurement of 2-port DUTs (scheme A)\\
$k_r$	& RF amplitude-to-voltage gain $k_r=v_o/\alpha_\text{RF}$ 
		    [Eq.~\req{eqn:amx-real-mixer-output-2}]\\
		& in the measurement of oscillators (schemes B, and C)\\
$k_{sd}$& LO amplitude-to-voltage gain $k_l=v_o/\alpha_\text{LO}$ 
		    [Eq.~\req{eqn:amx-real-mixer-output-3}]\\
		& in the bridge method (scheme D)\\
$\ell$	& mixer ssb voltage loss (thus, the ssb power loss is $\ell^2$)\\
$m$	& no.\ of averaged spectra\\
$P_0$         & carrier power.  Also $P_a$, $P_b$, $P_m$, etc.\\
$R_0$		  & characteristic resistance.  Often $R_0=50$ \ohm\\
rms		& root mean square value\\
$S(f)$, $S_x(f)$ & single-sided power spectrum density (of the quantity $x$)\\
$t$                & time\\
$T_0$	& absolute temperature, reference temperature ($T_0=290$ K)\\
$v(t)$		& (voltage) signal, as a funtion of time\\
$v_o(t)$	& mixer output voltage, as a funtion of time\\
$V_0$		& peak carrier voltage (not accounting for noise)\\
$x(t)$, $y(t)$	& mixer output voltage (in two-channel systems)\\
$X(f)$, $Y(f)$	& one-sided Fourier transform of $x(t)$ and $y(t)$\\
$\alpha(t)$        & fractional amplitude fluctuation\\
$\lambda$		& wavelength\\
$\nu_0$	& carrier frequency\\
\rule[-1ex]{0pt}{0ex}%
$\phi(t)$          & phase fluctuation\\\hline
	%\end{tabular}
\end{longtable}
\end{center}
\clearpage

\tableofcontents
\cleardoublepage

%---------------------------------------------------------------------
\section{Introduction}\label{sec:amx-introduction}
%---------------------------------------------------------------------
The phase noise of oscillators and of two-port devices is a relevant issue in time-and-frequency metrology, in experimental physics, in space exploration, and in some fields of electronics, which include at least instrumentation, telecommunications, high speed digital circuits, and radar systems. 

Let us introduce the quasi-perfect sinusoidal signal of frequency $\nu_0$
\begin{equation}
v_i(t)=V_0[1+\alpha(t)]\cos[2\pi\nu_0t+\phi(t)]~,
\label{eqn:amx-noisy-sinusoid}
\end{equation}
in which $\phi(t)$ and $\alpha(t)$ are the random phase fluctuation and the normalized random amplitude fluctuation, respectively.
Phase noise is usually described in term of $S_\phi(f)$, namely, the power spectral density (PSD) of $\phi(t)$ as a function of the Fourier frequency $f$.  Similarly, $S_\alpha(f)$ is the PSD of $\alpha(t)$. In practice, the PSD is measured as the average square modulus of the one-sided Fourier transform normalized for the power-type signals.  $S_\phi(f)$ is used to describe fast fluctuations, while time-domain measurements are preferred for slow fluctuations. The boundary is generally set at $10^{-2}$ to 1 Hz. 
The general background on phase noise and on frequency stability is available from numerous references, among which we prefer  \cite{chronos:frequency,kroupa:frequency-stability,ieee99std1139,ccir90rep580-3}.

Phase noise is measured by means of a phase detector followed by a low-noise dc amplifier and a fast Fourier transform (FFT) analyzer.  In most cases the detector is a saturated double-balanced mixer \cite{nelson04fcs,agilent:E5500-phase-noise,aeroflex:PN9000-phase-noise}.  A balanced bridge (often referred to as `interferometer') with amplification and synchronous detection of the noise sidebands is used when the highest sensitivity is required \cite{sann68mtt,labaar82microw}.  The sensitivity is limited by the equivalent temperature of the instrument \cite{ivanov98uffc}.  Improved sensitivity is obtained by correlation and averaging, with two separate---thus independent---systems that measure the same device under test (DUT) \cite{vessot64nasa,walls76fcs}.  The dual-bridge with correlation exhibits the highest reported sensitivity, limited by the thermal uniformity of the instrument instead of the absolute temperature \cite{rubiola00rsi-correlation}.

We observed that amplitude noise (AM noise), always present in the system, limits the sensitivity by breaking the hypothesis of statistical independence. The steeper is the spectrum slope ($1/f$ and $1/f^2$), more disturbing is the effect at low Fourier frequencies. 
Thus, we stress the importance of AM noise in the emerging domain microwave photonics \cite{chang:rf-photonics}, where the laser RIN has a spectrum $1/f^2$ at low frequencies.
Unfortunately, little information on AM noise is available (see Ref.~\cite{rubiola05arxiv-am-noise}).  A careful analysis of the correlation schemes is necessary to understand the effect of AM noise and when it can be reduced or eliminated. After that, the measurement of the detector parameters turns out to be surprisingly simple.

%---------------------------------------------------------------------
\section{Phase noise measurements}\label{sec:amx-schemes}
%---------------------------------------------------------------------
\begin{figure}[t]
\centering\namedgraphics{\normalsizescale}{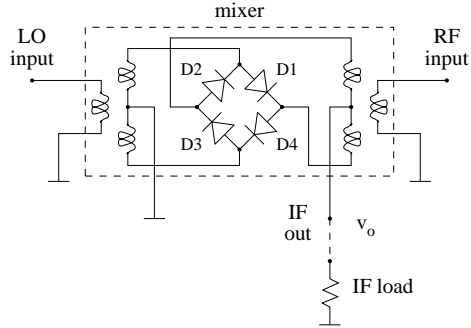}
\caption{Double balanced mixer.}
\label{fig:amx-dbm}
\end{figure}
Saturated by two signals of power of 3--30 mW (5--15 dBm) in quadrature with one another, the Schottky-diode double-balanced mixer (Fig.~\ref{fig:amx-dbm}) works as a phase detector governed by 
\begin{equation}
v_o(t)=k_\phi\phi(t)~.
\label{eqn:amx-ideal-mixer-output}
\end{equation}
The phase-to-voltage gain $k_\phi$ is an experimental coefficient that depends on technology and on power.  Actual values are of  0.1--0.5 V/rad.
It turns out that the background $1/f$ noise is chiefly due to the mixer (about $-120$ \unit{dBrad^2/Hz} for the microwave mixers, and about $-140$ \unit{dBrad^2/Hz} for the HF-UHF mixers).  Conversely, the background white noise comes from the dc preamplifier at the mixer output.  This is due to the low value of $k_\phi$, in conjunction with the technical difficulty of designing a dc amplifier noise-matched to the low output impedance (50 \ohm) of the mixer output. A floor of $-160$  \unit{dBrad^2/Hz} is common in practice.  

\begin{figure}[t]
\centering\namedgraphics{\normalsizescale}{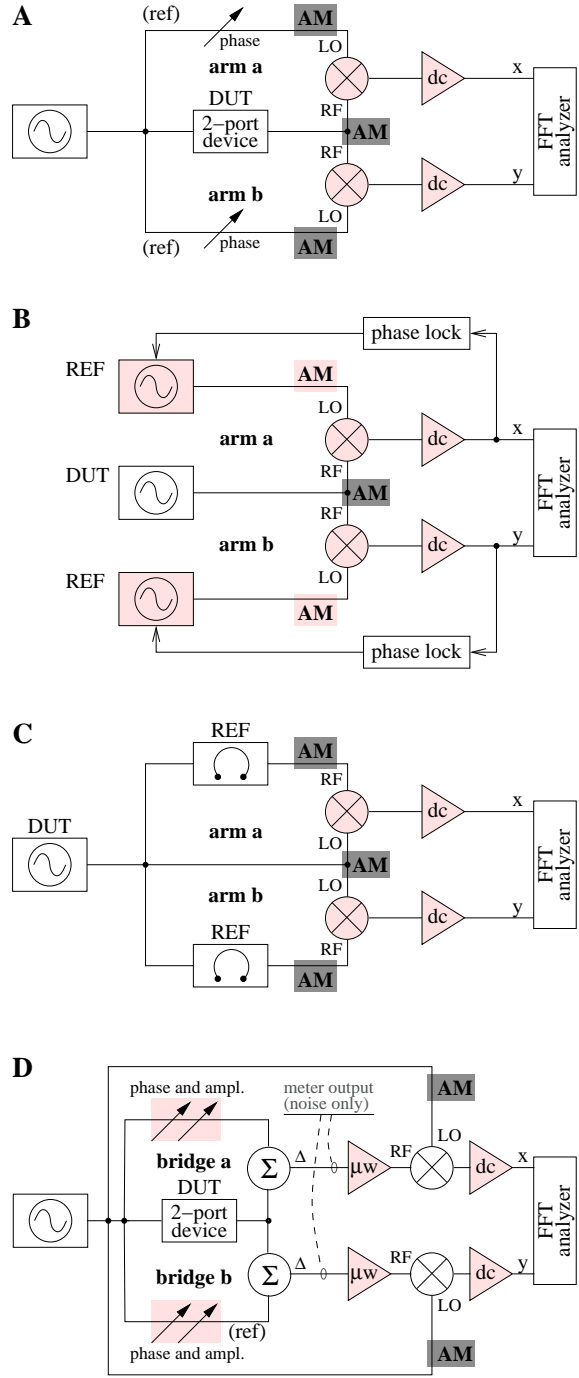}
\caption{Dual-channel (correlation) phase noise measurements. The noise sources rejected by correlation are colored in pink (light grey).  The AM noise not rejected by correlation is colored in dark grey.}
\label{fig:amx-dual-channel}
\end{figure}
Figure~\ref{fig:amx-dual-channel} shows the basic correlation schemes for the measurement of phase noise.  The light shadows indicate the sources of noise removed by correlation and averaging, while the dark shadows emphasize the points in which the effect of AM noise enters in the cross spectrum.

The scheme A is used to measure a two-port DUT \cite{walls76fcs}.
In order to reject the phase noise of the reference oscillator, the DUT group delay must be small. Phase adjustment is necessary to ensure the quadrature relationship.  Amplification or attenuation is needed if the DUT power does not fit the mixer input range.  Yet, the $1/f$ phase noise of the amplifier is generally higher than that of the mixer.

The scheme B serves to measure the phase noise of an oscillator.  This scheme is routinely used at the NIST for the measurement of low-noise oscillators using commercial synthesizers as the references \cite{nelson04fcs}.  A tight loop is advantageous vs.\ a loose loop \cite{audoin81ferro-milone} because it overrides the stray injection-locking, sometimes hardly avoidable, and because it relaxes the need for large dynamic range in the DAC converter of the FFT\@. Of course, the loop transfer function is to be measured accurately and taken away.

The scheme C makes use of two reference resonators that turns the oscillator frequency noise into phase noise at the mixer inputs.  The maximum frequency for the measurement of phase noise is limited by the resonator bandwidth.  Beyond, the  resonators attenuates the oscillator carrier, for the mixer is no longer saturated.
The reference resonator can be replaced with a delay line \cite{lance84}.  In this case, the maximum frequency is limited by the inverse delay.
A delay longer than 10--100 ns can only be obtained with a photonic delay line \cite{rubiola05josab-delay-line,salik04fcs-xhomodyne} because the loss of a coaxial cable is too high ($\sim1$ dB/m at 10 GHz for a 0.141-inch semirigid cable), while the optical fiber exhibits a loss of 0.2 dB/km (Corning SMF-28 at $\lambda=1.55$ $\mu$m). 
The single-channel version of the scheme C has been used to stabilize an oscillator either to a resonator \cite{galani84mtt} or to an optical-fiber delay line \cite{logan91fcs}.

In the scheme D, the mixer works in small-signal regime at the RF port, where only the amplified DUT noise is present.  The phase-to-voltage gain is 
\begin{math}
k_\phi = \frac{g}{2\ell}\sqrt{R_0P_0}
\end{math}, 
minus dissipative losses \cite{rubiola00rsi-correlation}.  $R_0$ is the characteristic resistance (50 \ohm), $P_0$ the DUT output power, $g$ is the voltage gain of the amplifier, and $\ell$ the ssb voltage loss of the mixer (in our early publications, $g$ and $\ell$ referred to power gain and loss). Thus, if $R_0=50$ \ohm, $P_0=10$ mW (10 dBm), $g=100$ (40 dB), and $\ell=2$ (the usual 6 dB loss of a mixer), the gain is $k_\phi\simeq17.7$ V/rad.
At the high sensitivity of the bridge, the $1/f$ noise of the variable phase shifter and attenuator shows up.  The single bridge can also be used to measure or to stabilize an oscillator \cite{ivanov98uffc}.

%---------------------------------------------------------------------
\section{The effect of AM noise on correlation systems}\label{sec:amx-correlation-am}
%---------------------------------------------------------------------
\begin{figure}
\centering\namedgraphics{\normalsizescale}{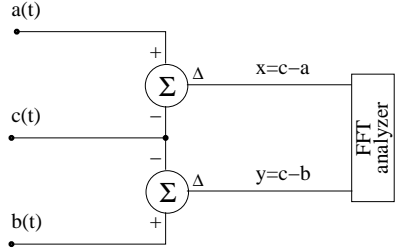}
\caption{Basic dual-channel (correlation) spectrum measurement.}
\label{fig:amx-correlation}
\end{figure}
Correlation works as shown in Fig.~\ref{fig:amx-correlation}, where the mixers are represented as $\sum$ nodes because they take the phase difference.  
Let us denote with $\mathbb{E}\left\{\,\right\}$ the statistical expectation, and with $\left<\,\right>_m$ the average on $m$ realizations.  
The expectation operator prevails over the average, thus $\mathbb{E}\left\{\left<\,\right>_m\right\}=\mathbb{E}\left\{\,\right\}$.
The power spectral densities are measured as
$\left<S_{xx}\right>_m=\left<XX^*\right>_m$ and 
$\left<S_{yy}\right>_m=\left<YY^*\right>_m$ for the single channel spectral density, and as
$\left<S_{yx}\right>_m=\left<YX^*\right>_m$ for the cross spectral density.
The uppercase $X$ and $Y$ are the one-sided Fourier transform of the lowercase variables, and the superscript `$*$' stands for complex conjugate.  

First, we assume that  $a$ and $b$ of  Fig.~\ref{fig:amx-correlation} are the statistically-independent single-channel background noises, and that $c$ is the DUT noise.  There follows that
\begin{align}
\mathbb{E}\left\{\left<S_{yx}\right>_m\right\} & = S_{cc}
\end{align}
because $\mathbb{E}\left\{S_{ba}\right\}=0$.  The DUT noise is measured in this way.

Then we set $c=0$, for $\mathbb{E}\left\{\left<S_{yx}\right>_m\right\}= 0$ holds.  This gives the background noise of the instrument as the variance
\begin{align}
\mathrm{VAR}\left\{\left<S_{yx}\right>_m\right\} 
& = \mathbb{E}\bigl\{\left|\left<S_{yx}\right>_m - \mathbb{E}\bigl\{S_{yx}\bigl\}\right|^2 \bigr\}\\
& = \mathbb{E}\bigl\{\left|\left<S_{yx}\right>_m\right|^2 \bigr\}~,
\end{align}
which is proportional to $1/m$.

Owing to the asymmetry of the diodes and of the baluns (transformers), the mixer (Fig.~\ref{fig:amx-dbm}) is not perfectly balanced.
Hence, the signal power affects the dc offset at the mixer output.  Consequently, the AM noise taken in in this way can not be rejected by correlation.
Nulling the sensitivity to AM noise is an issue, which can be tackled by playing on power and on the quadrature relationship.  This was reported long time ago with old HF mixers \cite{brendel75im}.  It was also suggested that the mixer can be set to a sweet point, off the quadrature condition, where the sensitivity to AM noise nulls, and the mixer is still a valuable phase detector.  A similar approach was followed in \cite{cibiel02uffc}, with microwave mixers.  
Yet, nothing is said about the generality of the method versus the mixer type, and versus the measurement scheme.  

First, we observed experimentally that the output of a saturated mixer is of the form
\begin{align}
\label{eqn:amx-real-mixer-output-1}
v_o(t) & = k_\phi\,\phi(t) + k_{lr}\,\alpha(t) 	&&\text{(scheme A)}
\intertext{if the two inputs see the same AM noise, of the form} 
\label{eqn:amx-real-mixer-output-2}
v_o(t) &= k_\phi\,\phi(t) + k_l\,\alpha_l(t) + k_r\,\alpha_r(t)  &&\text{(B and C)}
\intertext{if the two input see separate AM noises, and that Eq.~\req{eqn:amx-real-mixer-output-2} turns into}
\label{eqn:amx-real-mixer-output-3}
v_o(t) &= k_\phi\,\phi(t) + k_{sd}\,\alpha_l(t)		&&\text{(scheme D)}
\end{align}
for the bridge scheme, where only the LO port is saturated.  The subscripts $l$ and $r$ refer to LO and to RF, and $sd$ to synchronous detection.

The scheme A is a simple case.  As the AM noise is described by
\begin{align}
x(t) &= (k_{lr})_a \, \alpha(t)\\
y(t) &= (k_{lr})_b \, \alpha(t)~,
\end{align}
the two variable phases can be adjusted separately for the corresponding mixer to operate at the sweet point, if it exists.

The scheme B is unfortunate because the AM noise is governed by 
\begin{align}
x(t) &= (k_{r})_a \, \alpha(t)*h_a(t) + (k_{l})_a \, \alpha(t)\\
y(t) &= (k_{r})_b \, \alpha(t)*h_b(t) + (k_{l})_b \, \alpha(t)~.
\end{align}
The convolution ($*$) with the resonator low-pass transfer function $h$ de-correlates the DUT AM noise at inputs of the mixer by introducing the resonator group delay in one branch.  This de-correlation effect is inevitable because it is the same mechanism exploited to measure the DUT PM noise.
Of course, there is no way to null both $k_r$ and $k_l$ of the same mixer by playing with the phase around the quadrature.  Unless this occurs unexpectedly, out of good luck.  

The scheme C is ruled by 
\begin{align}
x(t) &= (k_{l})_a \, \alpha_a(t) + (k_{r})_a \, \alpha_\text{\textsc{dut}}(t)\\
y(t) &= (k_{l})_b \, \alpha_b(t) + (k_{r})_b \, \alpha_\text{\textsc{dut}}(t)~.
\end{align}
In this case, the AM noise of the reference oscillators is rejected by correlation because the two oscillators are independent.  It is therefore sufficient to null the two $k_r$.  
Of course the sweet point, if exists, is not the same as for the scheme A\@.
The off-quadrature phase is set by adding a dc term at the input of the phase-lock circuit.  A sharp null is found by inspecting on the mixer output with a lock-in amplifier, after modulating the DUT output. The amplitude modulator must have no residual phase modulation.  

\begin{figure}[t]
\centering\namedgraphics{\normalsizescale}{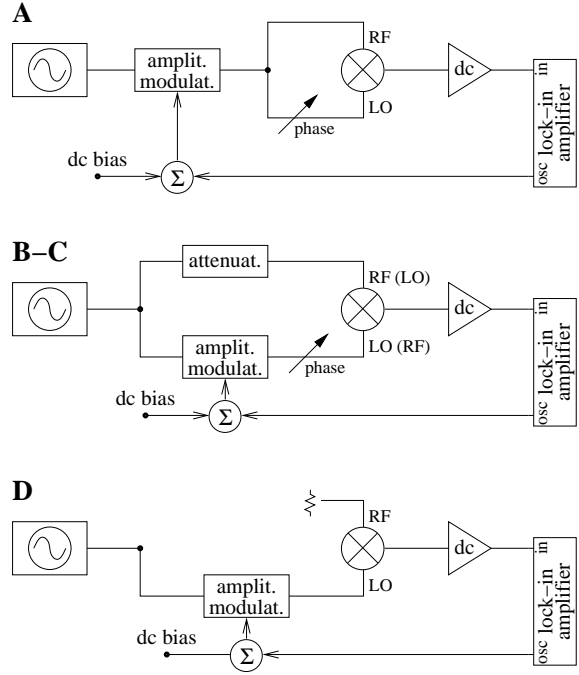}
\caption{Measurement of the mixer sensitivity to AM noise.}
\label{fig:amx-mixer-measurement}
\end{figure}

In the scheme D, the effect of the AM noise is 
\begin{align}
x(t) &= (k_{sd})_a \, \alpha(t)\\
y(t) &= (k_{sd})_b \, \alpha(t)~.
\end{align}
The need for AM noise rejection may depend on the microwave gain that precedes the mixer because this parameter influences the ratio $k_\phi/k_{sd}$.
Yet, even if there is only one parameter, it can not be nulled by offsetting the phase.  This occurs because the synchronous detection detects the DUT noise according to 
\begin{align}
x(t) & \propto \alpha(t)\sin\gamma + \phi(t)\cos\gamma~,
\end{align}
where $\gamma$ is the phase of the mixer LO signal. 
Consequently, $\gamma\neq0$ results in the DUT AM noise to be mistaken for PM noise.  
The solution, if any, comes from quite a different approach.  We are exploring a chopper technique, similar to the Dicke radiometer \cite{dicke46rsi-radiometer}.

\begin{figure}[t]
\centering\namedgraphics{\normalsizescale}{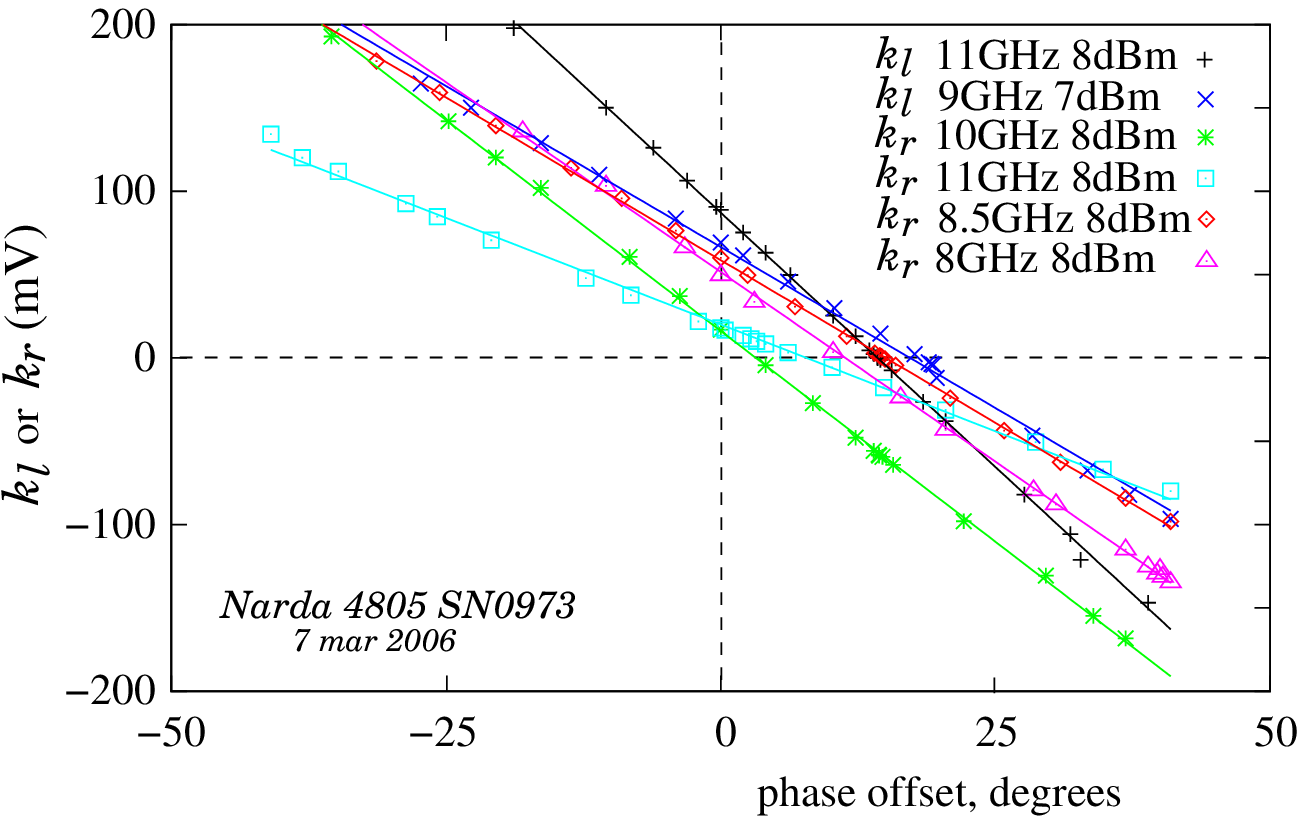}\\[1em]
\centering\namedgraphics{\normalsizescale}{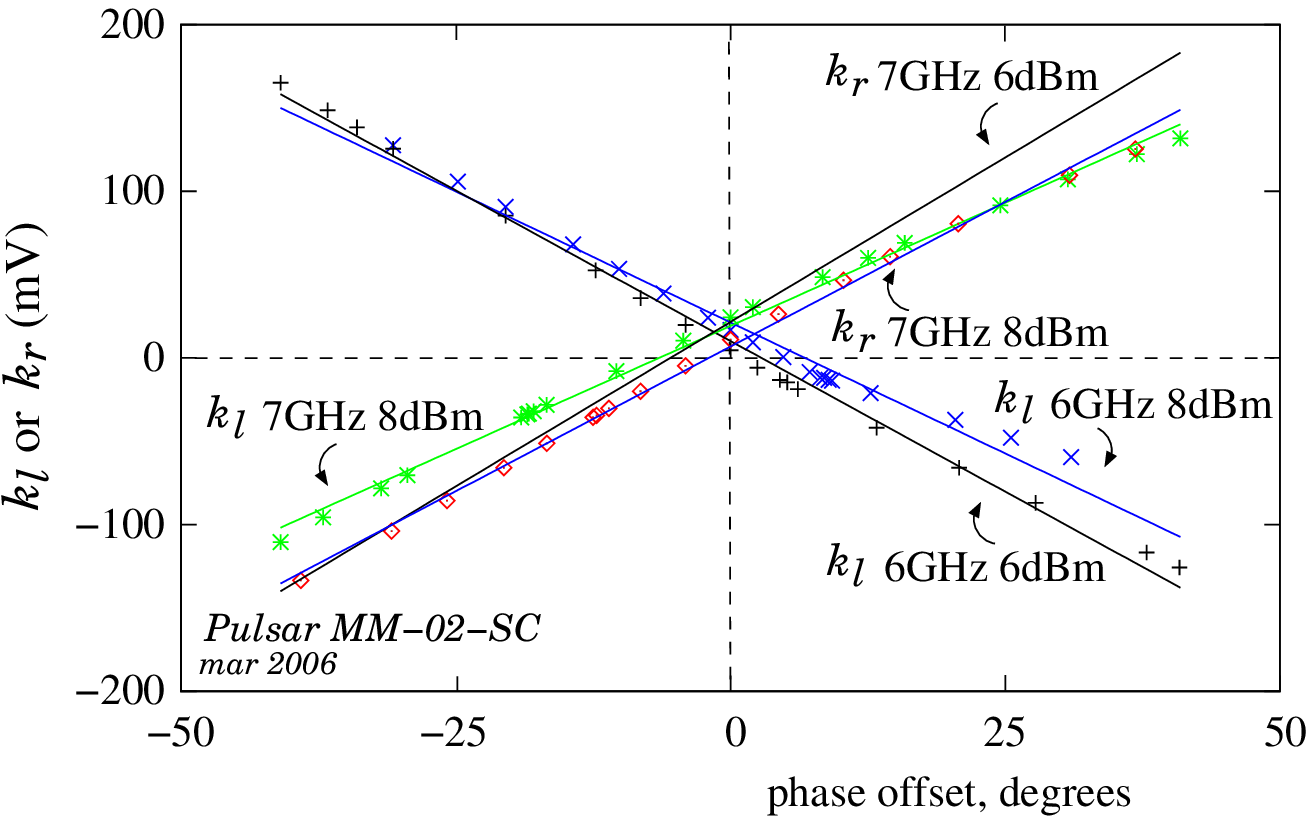}
\caption{AM sensitivity of two microwave mixers.}
\label{fig:amx-kalpha-vs-phase}
\end{figure}

%---------------------------------------------------------------------
\section{Mixer measurement}\label{sec:amx-mixer}
%---------------------------------------------------------------------
We validate our analysis with the experiments of Fig.~\ref{fig:amx-mixer-measurement},
which also provide the actual parameters of some mixers.  The mixers are selected among those routinely used in our laboratory, and tested in the same conditions as in the measurement of phase noise.  These mixers are not special devices for phase noise measurement.  Instead, they are high-performance general-purpose devices for microwave and radio engineering.

\subsection{Microwave mixers}
%---------------------------------------------------------------------
Out of experimental selection, we found an amplitude modulator that shows a null of residual phase modulation at a given dc bias, where the device also shows a sufficiently small attenuation (1.5 dB). This loss is compensated by changing the source power in Fig.~\ref{fig:amx-correlation}~A and D, and with an attenuator in Fig.~\ref{fig:amx-correlation}~B and C\@.  The modulator gain is $\alpha/v_\text{in}=7.2{\times}10^{-2}$ \unit{V^{-1}} (0.625 dB/V).  In order to avoid any nonlinear effect we set the microwave modulation to a low value, $\alpha_\text{rms}=7.2{\times}10^{-3}$, so that the mixer output never exceeds 350 \unit{\mu{V}_{rms}}.  The bandwidth of the modulation channel, from the ac input of the sum node to the output of the dc amplifier, is large (1 MHz, limited by the dc amplifier) as compared to the measurement frequency (10 kHz), thus there is no phase lag.  The lock-in amplifier is set for the measurement of the real part, so it keeps the sign.  In actual phase noise measurements, it is vital to understand that this setting detects the sweet point as a smooth zero crossing.  Conversely, the measurement of the modulus shows a sharp cusp, hard to identify properly.
The lock-in can be replaced with a dual-channel FFT analyzer, used to measure the real part of the voltage ratio.

We measured a few microwave mixers in saturated conditions, modulating the amplitude at one input, as in Fig.~\ref{fig:amx-mixer-measurement}~B-C\@.  An example of results is reported in Fig.~\ref{fig:amx-kalpha-vs-phase}.

\begin{figure}[t]
\centering
\includegraphics[scale=\TINYscale]{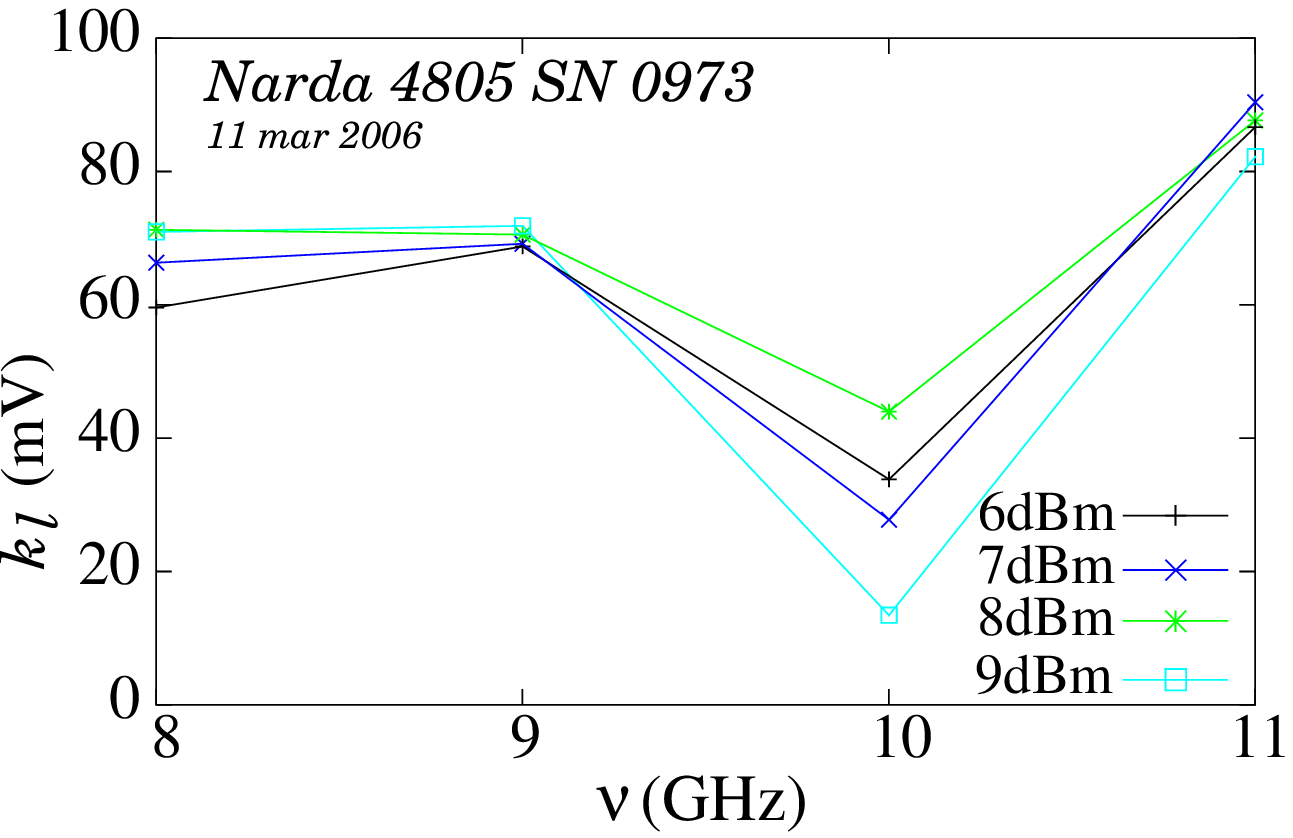}
\includegraphics[scale=\TINYscale]{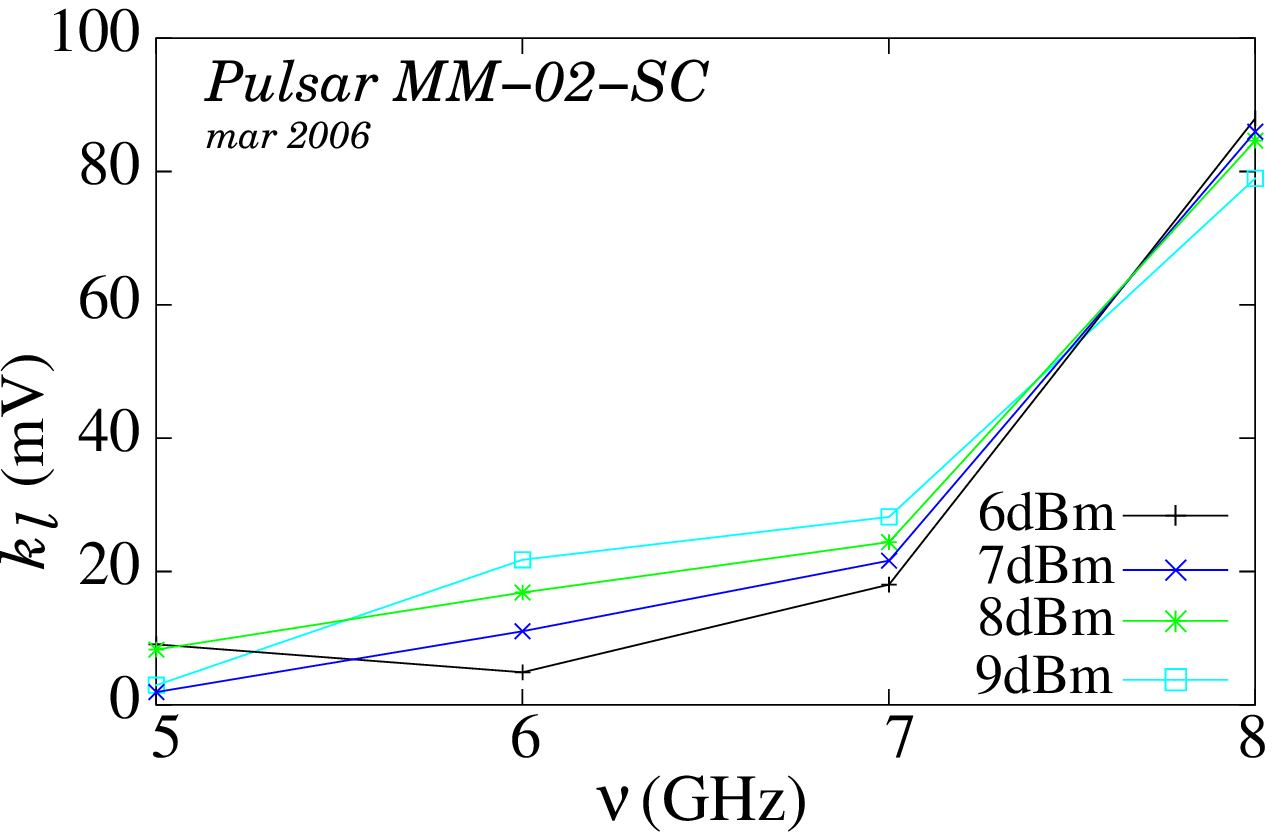}\\[1ex]
\includegraphics[scale=\TINYscale]{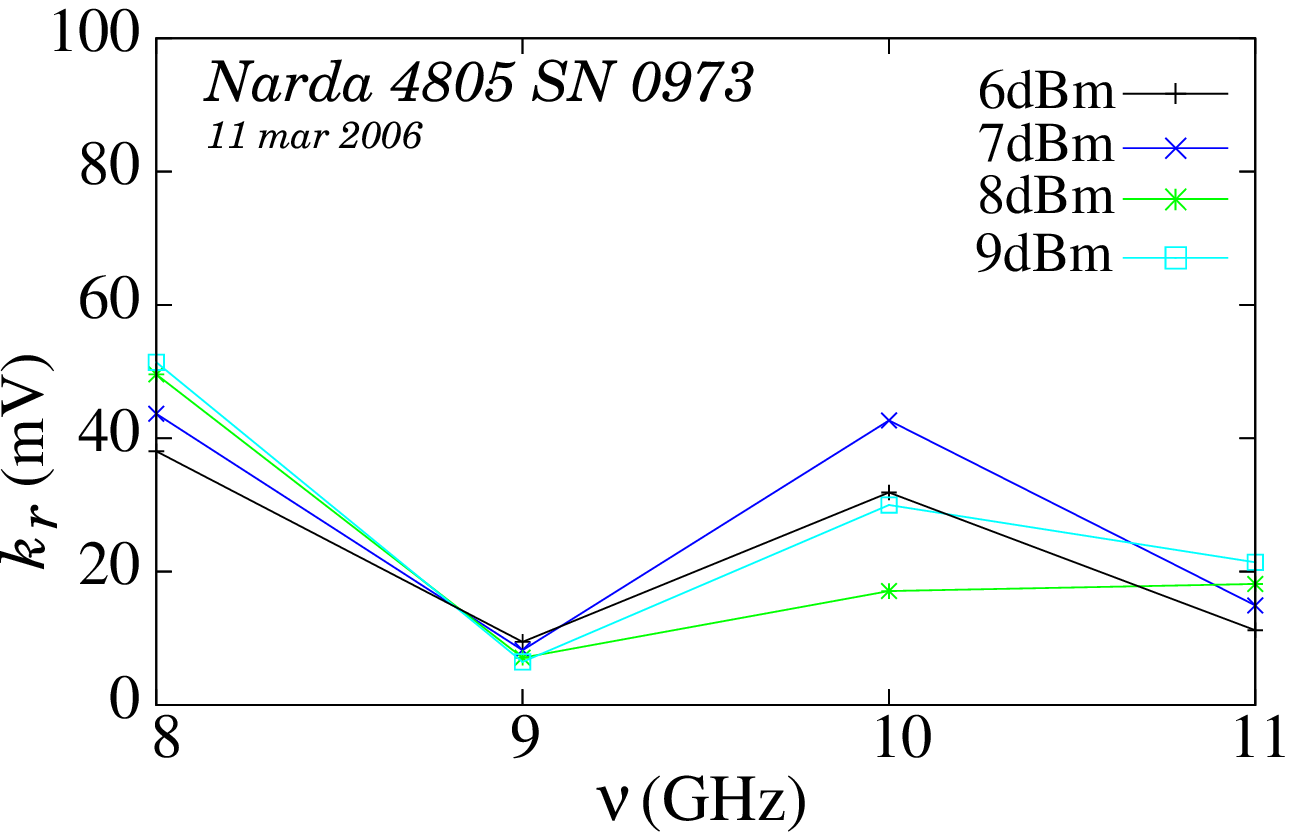}
\includegraphics[scale=\TINYscale]{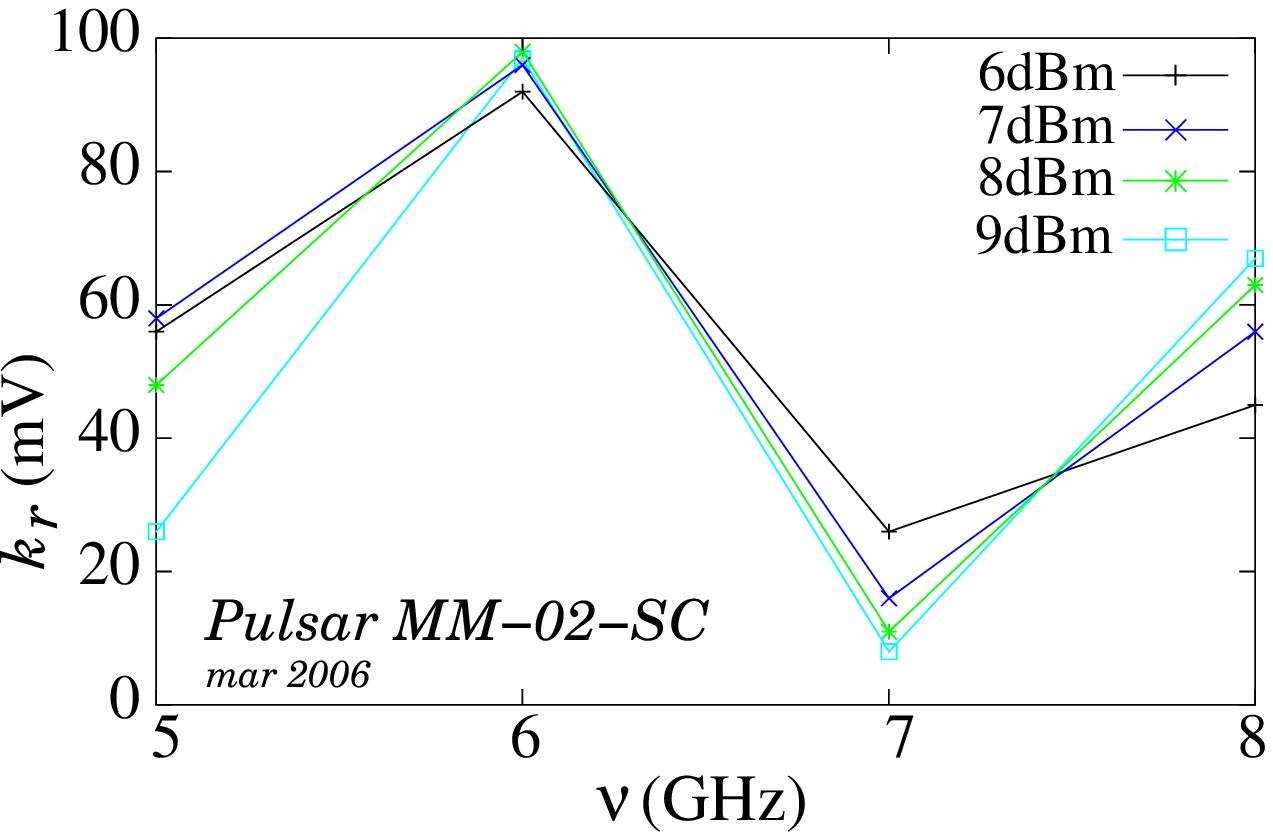}
\caption{Effect of power and frequency on $k_l$ and $k_r$ in some microwave mixers.}
\label{fig:amx-kalpha-vs-frequency}
\end{figure}

\begin{figure}
\centering
\includegraphics[scale=0.55]{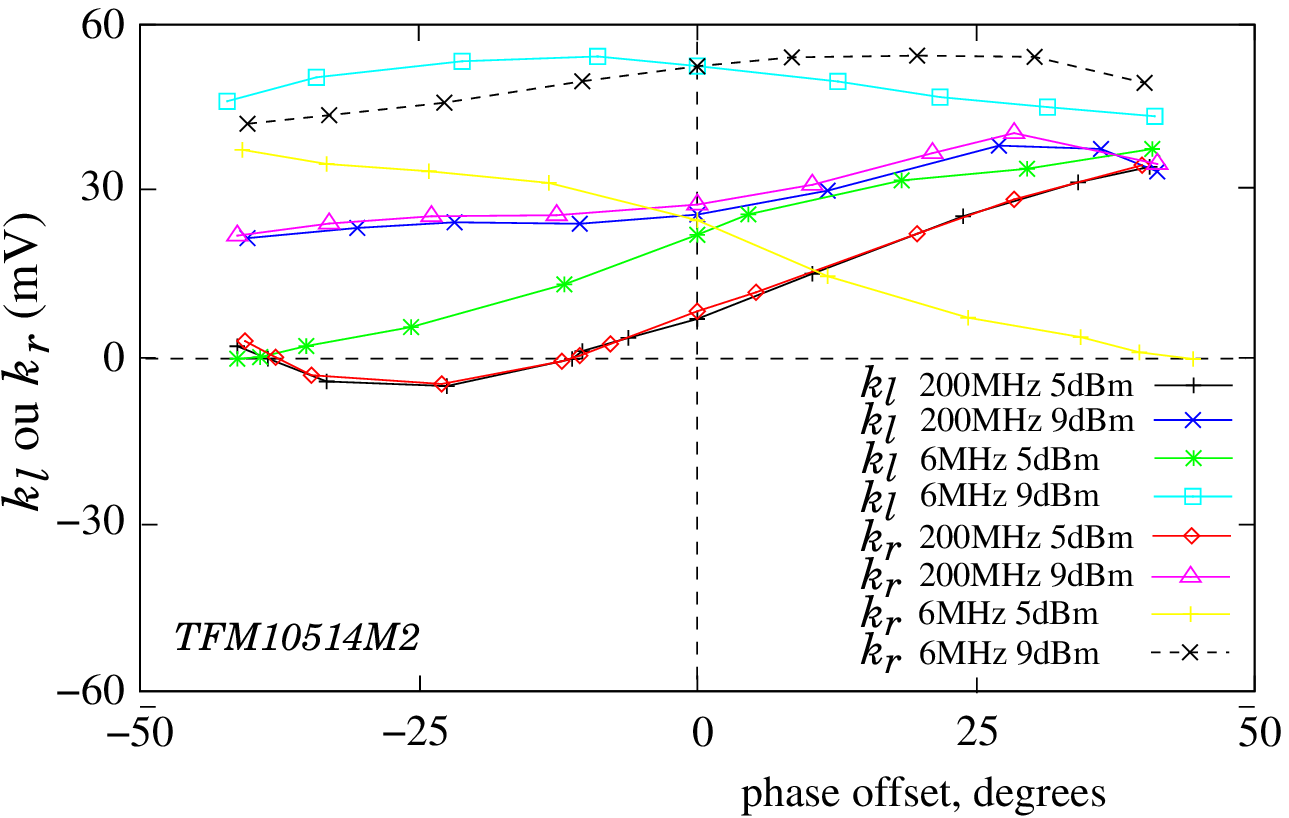}\\
\includegraphics[scale=0.55]{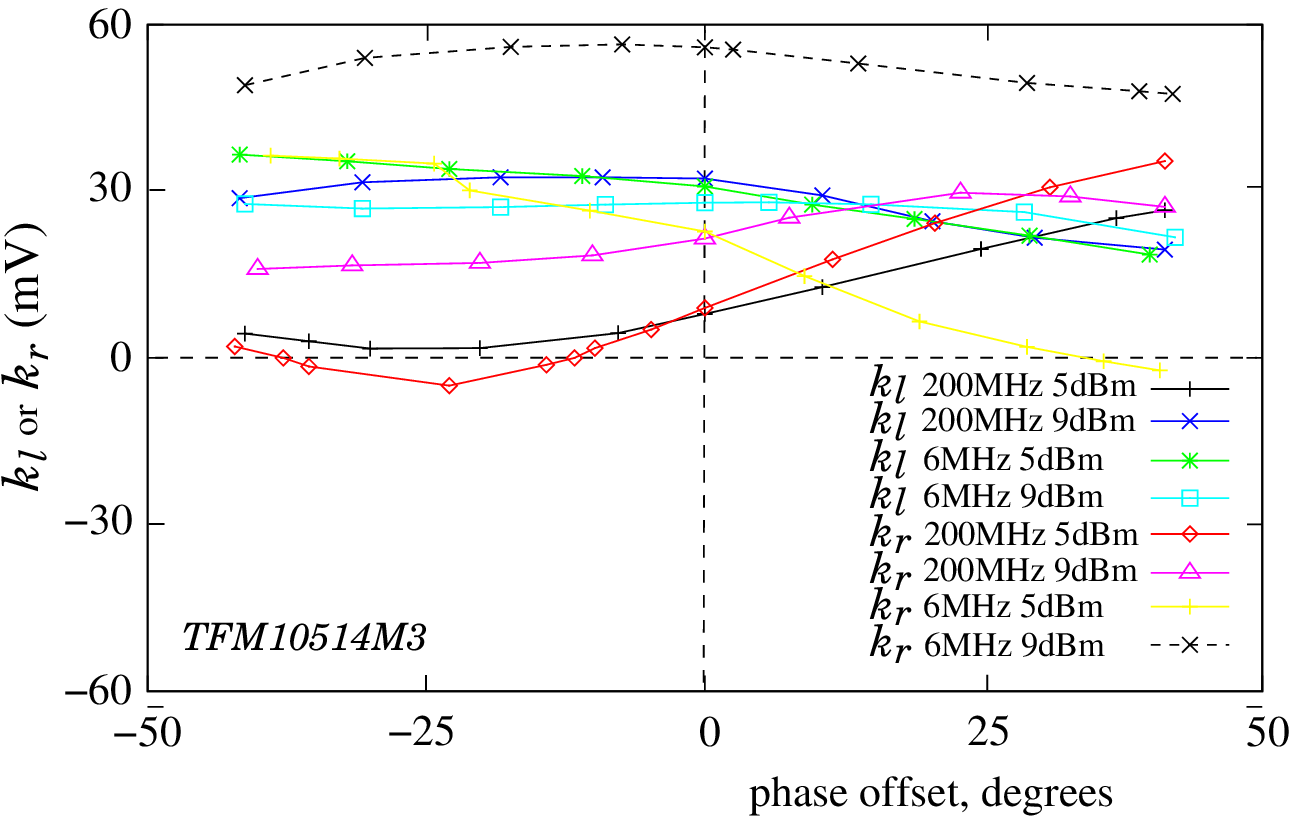}\\
\includegraphics[scale=0.55]{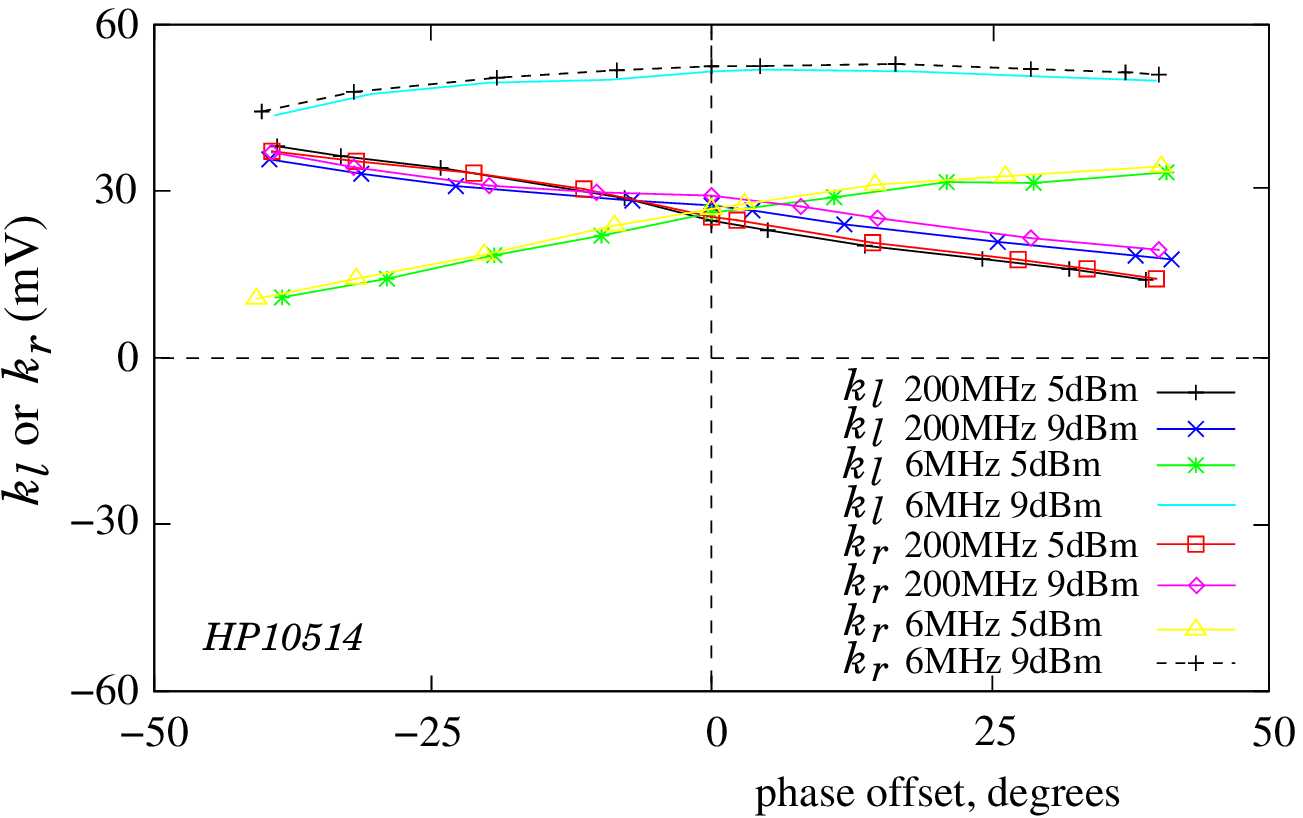}\\
\includegraphics[scale=0.55]{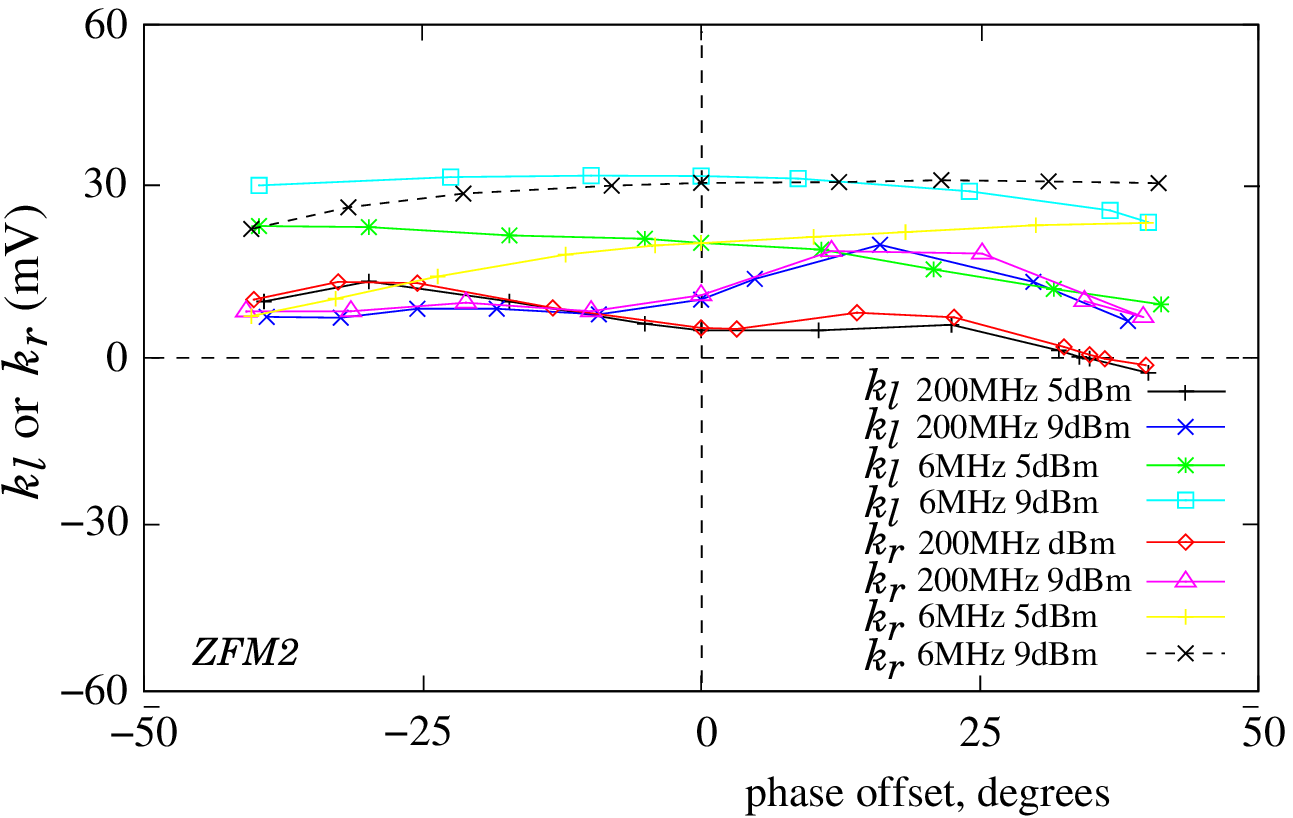}
\caption{Effect of power and frequency on $k_l$ and $k_r$ in some HF-VHF mixers.  For comparison, $k_\phi$ is of some 220 mV/rad.}
\label{fig:amx-vhf-mixers}
\end{figure}

With the Narda mixer, $k_l$ and $k_r$ show similar slope at different frequencies and power.  The curves are shifted towards right as frequency increases.  This makes one think to a systematic phase shift in the baluns.  In fact, the inside of a microwave mixer differs from Fig.~\ref{fig:amx-dbm} in that the transformers are replaced with microstrip networks.  In all the reported conditions, the null of AM sensitivity is clearly visible.  Yet it occurs at a phase up to $20^\circ$ off the quadrature, which may be too large.  
A lower $k_\phi$ at this large phase offset is only a minor problem.  The main problem is that the dc output voltage (100 mV) is too large for the precision dc amplifier that follows.  In fact, a gain of 40 dB or more is often needed to override the input noise of commercial FFTs.   

In the case of the Pulsar mixer, we observe that $k_l$ and $k_r$ can change sign unexpectedly with frequency, and that power has a minor effect.  On the other hand, the nulls are well clustered in a region of $\pm5^\circ$ around the quadrature, where the output voltage is within 30 mV\@.

The following Table shows all the $k$ coefficients for four mixers measured at 10 GHz.  
\begin{center}
\begin{tabular}{lccccc}
\multicolumn{6}{l}{\vspace*{-1ex}}\\\hline
\rule[-1ex]{0pt}{3ex}%
Mixer					&$k_\phi$ & $k_{lr}$ & $k_r$   & $k_l$ & $k_{sd}$\\\hline
\rule[0ex]{0pt}{2ex}%
Narda 4805 s.no.\,0972	& 272 & 16    &   7.9  & 37   & 6.5   \\
Narda 4805 s.no.\,0973	& 274 & 18.3 & 17.1 & 44    & 9.8   \\
NEL 20814				& 279 & 51.5 & 12.1 & 37.9 & 2.7   \\
\rule[-1ex]{0pt}{2ex}%
NEL 20814				& 305 & 41    & 1.9   & 30.2 & 3.73 \\\hline
\rule[-1ex]{0pt}{3ex}%
unit						&mV/rad& mV & mV & mV  & mV\\\hline
\multicolumn{6}{l}{\rule[0ex]{0pt}{2ex}%
Test parameters: $\nu_0=10$ GHz, $P=6.3$ mW (8 dBm)}
\end{tabular}
\end{center}
The variable phase was set for the output to be 0 V dc.  We observe that $k_{sd}$ is significantly different from $k_l$.  This is related to the fact that $k_l$ is measured with the RF port is saturated, while $k_{sd}$ is measured with the RF terminated.   Additionally, we notice that $k_{lr}$ differs significantly from $k_l+k_r$.  This is the signature of a bizarre saturated interaction, which indicates that there is no way to forecast a result by adding separate effects.

Figure~\ref{fig:amx-kalpha-vs-frequency} shows the effect of power and frequency on two mixers.  Most of the change in the AM sensitivity is due to frequency.  The same fact was observed on other devices, not reported here.  This reinforces the idea of  systematic phase errors in the baluns.
Understanding this effect is difficult because the literature is old (see for example \cite{kollberg:mixers,maas:mixers}), and the actual design is confidential.  Nonetheless, there is a simple physical interpretation.  Common sense suggests that the baluns are designed for the lowest power change in the desired frequency range.  In practice, this is close to the condition of maximally flat amplitude as a function of frequency.  The amplitude vs.\ phase relationship is governed by the Cauchy-Riemann condition for the uniqueness of the derivative in analytic functions.  
Accordingly, the phase vs.\ frequency function has the steepest slope where the amplitude vs.\ frequency function is flat.

\begin{figure}[t]
\centering
\includegraphics[scale=0.75]{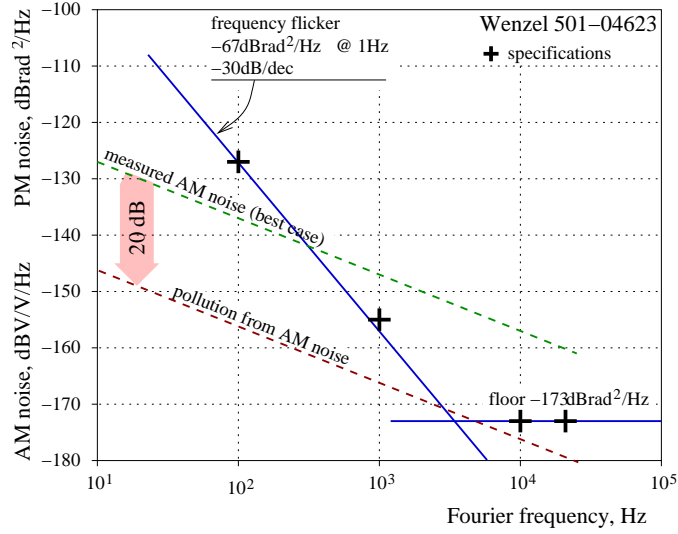}
\caption{The oscillator AM noise can pollute the phase noise measurement.}
\label{fig:amx-mywenzel}
\end{figure}

\subsection{HF-VHF mixers}
%---------------------------------------------------------------------
The selected mixers are suitable to the frequency range of 1--500 MHz, limited by the toroidal  transformers.  This range is typical for such devices.  
Driven at some 5--6 mW (7--8 dBm), the phase-to-voltage gain $k_\phi$ is of about 220 mV/rad, some 20\% lower than that of microwave mixers. 
The general behavior is rather uniform over the bandwidth, for there is no point in sweeping the frequency in small steps.  Thus, we choose two frequencies, 6 MHz and 200 MHz, determined by a specific application \cite{boudot06ell-x-band-oscillator}, and close enough to the frequencies of general interest (5, 10, 100 MHz).
The measurement system differs slightly from Fig~\ref{fig:amx-mixer-measurement}~B and C\@. We used two synthesizers driven by the same frequency standard, one adjusted in phase and the other modulated in amplitude with $\alpha=10^{-2}$ by a 1 kHz signal from the lock-in amplifier.
We focused on the schemes B-C\@.  The results are shown in Fig.~\ref{fig:amx-vhf-mixers}.  Surprisingly, in most cases there is no sweet point of zero sensitivity to AM\@.  The sweet point is present only in some specific conditions of power and frequency.  Yet, it appears at a large phase shift, up to $40^\circ$, where $k_\phi$ drops and the mixer is no longer usable as a phase detector. Besides, the large dc offset (up to 150 mV) makes the dc amplifier problematic.  Qualitative inspection on some other mixers confirms that this behavior is rather general.   

\section{Final remarks}
In the measurement of an oscillator the rejection of AM noise relies only on the mixer.  The AM noise of the reference can be rejected by correlation if two independent references are used (Fig.~\ref{fig:amx-dual-channel}~C).  Yet, correlation provides no rejection of the AM noise of the oscillator under test. 
The effect can be surprisingly high.  Figure~\ref{fig:amx-mywenzel} shows phase and amplitude noise of an ultra-stable quartz oscillator.  Phase noise comes from the manufacturer specifications, while the $1/f$ amplitude noise (taken from \cite{rubiola05arxiv-am-noise}) is the \emph{lowest} measured.   If the mixer's AM rejection ($k_\phi/k_l$, $k_\phi/k_r$, or $k_\phi/k_{lr}$)  is lower than some 20 dB, an experimental error shows up in the region of 3 kHz.  Of course, the mixer rejection can be significantly lower than 20 dB.

%==================================
\def\bibfile#1{/Users/rubiola/Documents/work/bib/#1}
%\def\bibfile#1{/home/rubiola/docs/bib/#1}
%==================================
\addcontentsline{toc}{section}{References}
\bibliographystyle{amsalpha}
\bibliography{\bibfile{ref-short},%
              \bibfile{references},%
              \bibfile{rubiola}}

\end{document}